\documentclass[aps,reprint,floatfix,superscriptaddress]{revtex4-1}
\usepackage{times}
\usepackage{graphicx}
\usepackage{soul}
\usepackage{siunitx} 
\usepackage[draft]{todonotes}   
\DeclareSIUnit \Ge{Ge}
\DeclareSIPostPower \pluspower{+}
\DeclareSIUnit \Mcps{Mcps}

\begin{document}

\title{
Quantum Nonlinear Optics with a Germanium-Vacancy Color Center \\ in a Nanoscale Diamond Waveguide
}
\author{M. K. Bhaskar}
\email{mbhaskar@g.harvard.edu}
\affiliation{Department of Physics, Harvard University, 17 Oxford Street, Cambridge, Massachusetts 02138, USA}
\author{D. D. Sukachev}
\affiliation{Department of Physics, Harvard University, 17 Oxford Street, Cambridge, Massachusetts 02138, USA}
\affiliation{P. N. Lebedev Physical Institute of the RAS, Moscow 119991, Russia}
\author{A. Sipahigil}
\affiliation{Department of Physics, Harvard University, 17 Oxford Street, Cambridge, Massachusetts 02138, USA}
\author{R. E. Evans}
\affiliation{Department of Physics, Harvard University, 17 Oxford Street, Cambridge, Massachusetts 02138, USA}
\author{M. J. Burek}
\affiliation{John A. Paulson School of Engineering and Applied Sciences, Harvard University, 29 Oxford Street, Cambridge, Massachusetts 02138, USA}
\author{C. T. Nguyen}
\affiliation{Department of Physics, Harvard University, 17 Oxford Street, Cambridge, Massachusetts 02138, USA}
\author{\\ L. J. Rogers}
\affiliation{Institute for Quantum Optics, University Ulm, Albert-Einstein-Allee 11, 89081 Ulm, Germany}
\author{P. Siyushev}
\affiliation{Institute for Quantum Optics, University Ulm, Albert-Einstein-Allee 11, 89081 Ulm, Germany}
\author{M. H. Metsch}
\affiliation{Institute for Quantum Optics, University Ulm, Albert-Einstein-Allee 11, 89081 Ulm, Germany}
\author{H. Park}
\affiliation{Department of Chemistry and Chemical Biology, Harvard University, 12 Oxford St., Cambridge, Massachusetts 02138, USA}
\author{F. Jelezko}
\affiliation{Institute for Quantum Optics, University Ulm, Albert-Einstein-Allee 11, 89081 Ulm, Germany}
\author{M. Lon\v{c}ar}
\affiliation{John A. Paulson School of Engineering and Applied Sciences, Harvard University, 29 Oxford Street, Cambridge, Massachusetts 02138, USA}
\author{M. D. Lukin}
\email{lukin@physics.harvard.edu}
\affiliation{Department of Physics, Harvard University, 17 Oxford Street, Cambridge, Massachusetts 02138, USA}
\begin{abstract}
We demonstrate a quantum nanophotonics platform based on
germanium-vacancy (GeV) color centers in fiber-coupled diamond nanophotonic waveguides. 
We show that GeV optical transitions have a high quantum efficiency and are nearly lifetime-broadened in such nanophotonic structures.
These properties yield an efficient interface between waveguide photons and a single GeV without the use of a cavity or slow-light waveguide.
As a result, a single GeV center reduces waveguide transmission by $18 \pm 1\%$ on resonance in a single pass. 
We use a nanophotonic interferometer to perform homodyne detection of GeV resonance fluorescence.
By probing the photon statistics of the output field, we demonstrate that the GeV-waveguide system is nonlinear at the single-photon level.
\end{abstract}

\maketitle

\begin{figure}
	\includegraphics[width=\linewidth]{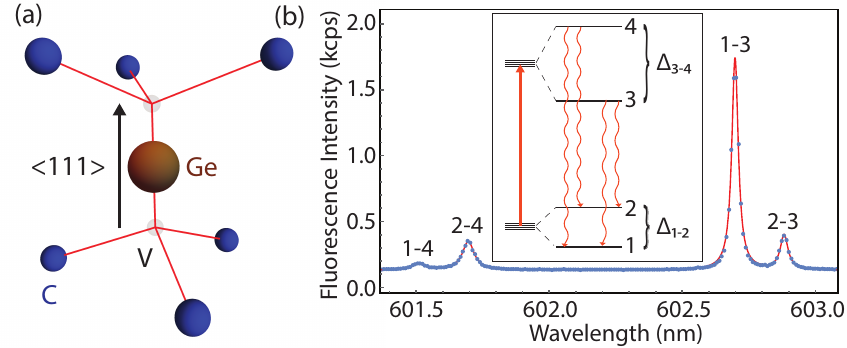}
		\caption{\label{fig1}
	(a) Molecular structure of the GeV center.
	 (b) Photoluminescence spectrum of the GeV at $T =$ \SI{50}{\K}, revealing the four optical transitions predicted by the GeV electronic structure (inset) \cite{palyanov2015germanium}. $\Delta_{1-2} = $ \SI{152}{\GHz} and $\Delta_{3-4} = $ \SI{981}{\GHz} are the measured ground and excited state orbital splittings respectively. The solid curve is a fit to four Lorentzians.
	}
\end{figure}

Efficient coupling between single photons and coherent quantum emitters is a central element of quantum nonlinear optical systems and quantum networks \cite{kimble2008quantum, lodahl2015interfacing, carusotto2013quantum}.
Several atom-like defects in the solid-state are currently being explored as promising candidates for the realization of such systems \cite{aharonovich2016solid},
including the nitrogen-vacancy (NV) center in diamond, renowned for its long spin coherence at room temperature \cite{maurer2012room};
and the silicon-vacancy (SiV) center in diamond, which has recently been shown to have strong, coherent optical transitions in nanostructures \cite{evans2016narrow, sipahigil2016integrated, jantzen2016nanodiamonds,li2016nonblinking}.
The remarkable optical properties of the SiV center arise from its inversion symmetry \cite{hepp2014electronic},
which results in a vanishing permanent electric dipole moment for SiV orbital states, dramatically reducing their response to charge fluctuations in the local environment.
A large family of color centers in diamond are predicted to have inversion symmetry \cite{goss2005vacancy} and therefore may be expected to have similarly favorable optical properties.  
In this Letter, we demonstrate an efficient optical interface using negatively-charged germanium-vacancy (GeV) color centers integrated into nanophotonic devices with optical properties that are superior to those of both NV and SiV centers. These properties result in high interaction probabilities between individual GeV centers and photons in a single-pass configuration, even without the use of cavities or other advanced photonic structures. 
The GeV center is a new optically active color center in diamond \cite{palyanov2015germanium,iwasaki2015germanium,ekimov2015germanium,palyanov2016high}.
Its calculated structure, shown in Fig.~\ref{fig1}(a), is similar to that of the SiV center with D$_\textnormal{3d}$ symmetry \cite{goss2005vacancy, iwasaki2015germanium}.
The Ge impurity occupies an interstitial site between two vacancies along the $\langle 111 \rangle$ lattice direction \cite{iwasaki2015germanium, ekimov2015germanium, goss2005vacancy}, resulting in inversion symmetry. 
Fig.~\ref{fig1}(b) depicts the electronic level structure of the GeV \cite{iwasaki2015germanium, ekimov2015germanium} with a zero-phonon line (ZPL) transition at \SI{602}{\nm} which constitutes about $60\%$ of the total emission spectrum \cite{palyanov2015germanium}. Similar to 
the negatively-charged SiV center \cite{hepp2014electronic, muller2014optical}, the ground state of the GeV center is a spin-doublet ($S = 1/2$) \cite{siyushev2016optical} with double orbital degeneracy.
The ground and excited orbital states of the GeV are split by spin-orbit coupling, forming a four-level system visible in its cryogenic photoluminescence (PL) spectrum [Fig.~\ref{fig1}(b)] \cite{palyanov2015germanium, ekimov2015germanium,palyanov2016high}.
As demonstrated in the complementary Letter \cite{siyushev2016optical}, this electronic structure allows one to directly control both orbital and electronic spin degrees of freedom using optical and microwave fields.

%

\begin{figure*}
	\includegraphics{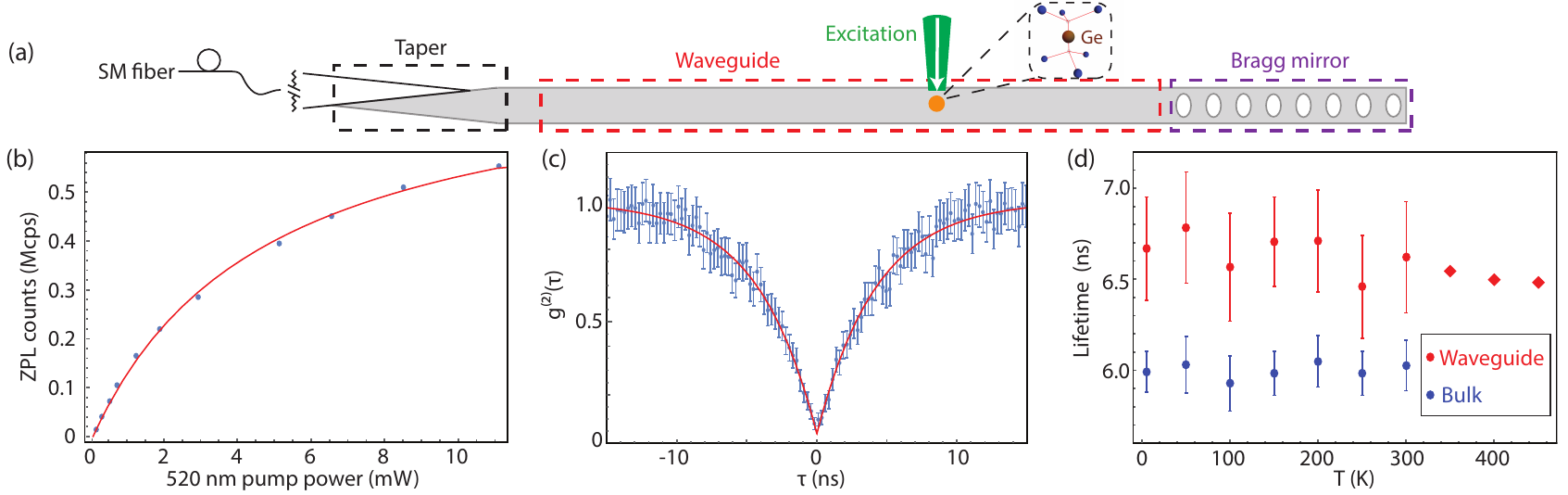}
		\caption{\label{fig2}
	(a) Schematic of a diamond nanophotonic device. Devices are \SI{100}{\micro\meter} long and \SI{480}{\nano\meter} wide and consist of a waveguide (red box), a partially-reflective Bragg mirror (purple box), and a taper (black box) for coupling to a tapered single-mode optical fiber.
	(b) Saturation response for a single GeV under continuous-wave, \SI{520}{\nm} excitation at $T = $ \SI{300}{\K}, measured as a function of optical power at the microscope objective. The solid curve is a fit to a two-level saturation model \cite{SOM}.
	(c) Intensity autocorrelation demonstrates antibunching of $g^{(2)} ( 0 ) = 0.08\pm 0.02$.
	The solid curve is a single-exponential fit.
	(d) GeV excited state lifetime measurement at different temperatures in waveguides (red) and bulk diamond (blue). Error bars represent standard deviation of measured lifetimes of seven different emitters. For $T>$ \SI{300}{\K}, the lifetime was measured for a single GeV in a waveguide (red diamonds).
}
\end{figure*}

We achieve efficient coupling of individual GeV centers with single photons by incorporating them into one-dimensional waveguides [Fig.~\ref{fig2}(a)] with transverse dimensions on the order of the single-atom scattering cross-section \cite{chang2007single,goban2015superradiance,javadi2015single, sipahigil2016integrated}.
Waveguides have a width of \SI{480}{\nm}, and are nanofabricated from diamond \cite{burek2014high, burek2016fiber}.
GeV centers are incorporated into devices at low density using $^{74}$\si{\Ge \pluspower} ion implantation ($10^9$ \si{\Ge \pluspower \per \square \cm})
and subsequent high temperature annealing at $\SI{1200}{\celsius}$, leading to spatially-resolvable single emitters \cite{evans2016narrow, iwasaki2015germanium}.
We couple a single-mode tapered optical fiber to the waveguide with $\sim 50\%$ coupling efficiency by positioning it in contact with a tapered section of the diamond
\cite{burek2016fiber, tiecke2015efficient, sipahigil2016integrated}.
We implement this technique
under a confocal microscope (described in \cite{sipahigil2016integrated, SOM}), enabling both free-space and fiber-based collection of fluorescence from GeV centers.

We first measure the ZPL emission of a single GeV in a waveguide under continuous-wave \SI{520}{\nm} off-resonant excitation at room temperature [Fig. \ref{fig2}(b)].
Collecting via the tapered fiber, we observe a maximum single-photon detection rate of $0.56 \pm 0.02$ \si{\Mcps} (million counts per second) on the narrowband ZPL around \SI{602}{\nm}, limited by excitation laser power.
The single-photon nature of the emission is verified by antibunching of ZPL photons [Fig. \ref{fig2}(c)], measured at $I/I_{sat} \sim 0.5$ where $I$ and $I_{sat}$ ($4.7 \pm 0.2$ \si{\mW} at \SI{520}{\nano \meter}) are applied and saturation intensities respectively.
To better understand the optical properties of the GeV, we measure its excited state lifetime at different temperatures with \SI{532}{\nm} pulsed excitation [Fig. \ref{fig2}(d)]. The GeV lifetime does not display significant temperature dependence up to $T = $ \SI{450}{\K} at which local vibrational modes with $\sim $ \SI{60}{\meV} energy  \cite{palyanov2015germanium,ekimov2015germanium} have finite occupation ($\bar{n} \sim 0.27$).
This demonstrates that multi-phonon relaxation paths play a negligible role in determining the excited state lifetime \cite{jahnke2015electron}, suggesting a high radiative quantum efficiency.
A statistically significant difference in lifetime for waveguide ($6.6 \pm 0.3$ \si{\ns}) and bulk ($6.0 \pm 0.1$ \si{\ns}) emitters implies a high sensitivity to the local photonic density of states \cite{khalid2015lifetime, frimmer2013nanomechanical, patel2016efficient}, providing further evidence of a high radiative quantum efficiency.
%

\begin{figure}
	\includegraphics[width=\linewidth]{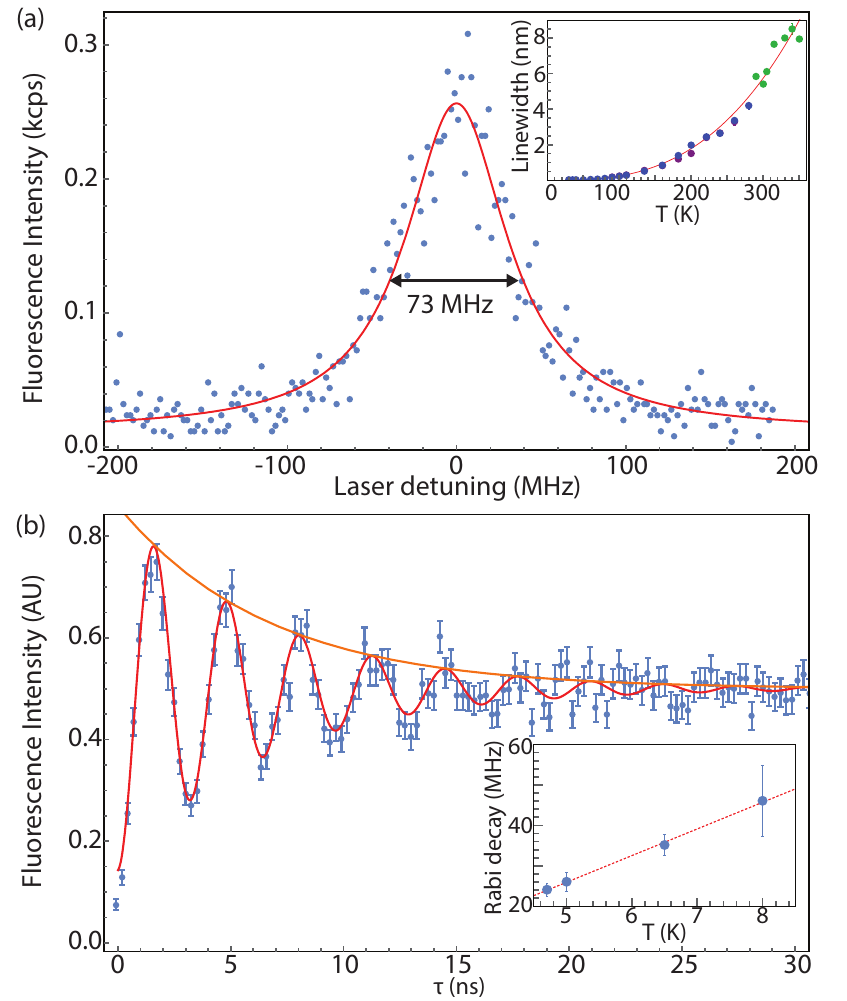}
		\caption{\label{fig3}
	(a) Transition 1-3 linewidth of a GeV in a waveguide at $T = $ \SI{5}{\K}, taken under resonant excitation at $I/I_{sat} \sim 0.01$. The solid curve is a Lorentzian fit. (Inset) Linewidth as a function of temperature, PL spectrum measured on a spectrometer under \SI{520}{\nm} excitation. Different colored points correspond to different emitters. The solid curve is a fit to a $T^3$ model \cite{SOM}.
	(b) Optical Rabi oscillations. Fluorescence is measured on the PSB under resonant excitation. The solid curves are fits to a two-level model \cite{SOM}. (Inset) The Rabi oscillation decay rate scales linearly as a function of temperature for $T < $ \SI{10}{\K}.
	}
\end{figure}

To study coherence properties of single GeV centers in a waveguide,  we use resonant excitation on transition 1-3 [Fig. 1(b)] at $T = $ \SI{5}{\K}. We scan the frequency of the laser over the GeV resonance and record the fluorescence in the phonon-sideband (PSB) collected into the tapered fiber.
This technique yields a linewidth of $73 \pm 1$ \si{\MHz} after $5$ minutes of averaging at low excitation intensity [Fig. \ref{fig3}(a)].
The measured linewidth of a GeV in a nanophotonic structure is within a factor of 3 of the lifetime-broadened limit, $\gamma_0/(2\pi) = 26\pm 1$ \si{\MHz} \cite{SOM}.
The measured linewidth increases with temperature up to $T =$ \SI{300}{\K} [inset of Fig. \ref{fig3}(a)], due to phonon broadening that scales as $a + b (T - T_0)^3$ [$a = 0.0 \pm 0.2$ \si{\nm}, $b = (1.9 \pm 0.5) \times 10^{-7}$ \si{\nano \meter \per \K \cubed}, $T_0 = (-13 \pm 23)$ \si{\K}] for $T > $ \SI{50}{\K}. The $T^3$ scaling suggests that optical coherence is limited by a two-phonon orbital relaxation process for $T > $ \SI{50}{\K}, similar to the case of the SiV \cite{jahnke2015electron}. 

\begin{figure}
	\includegraphics[width=\linewidth]{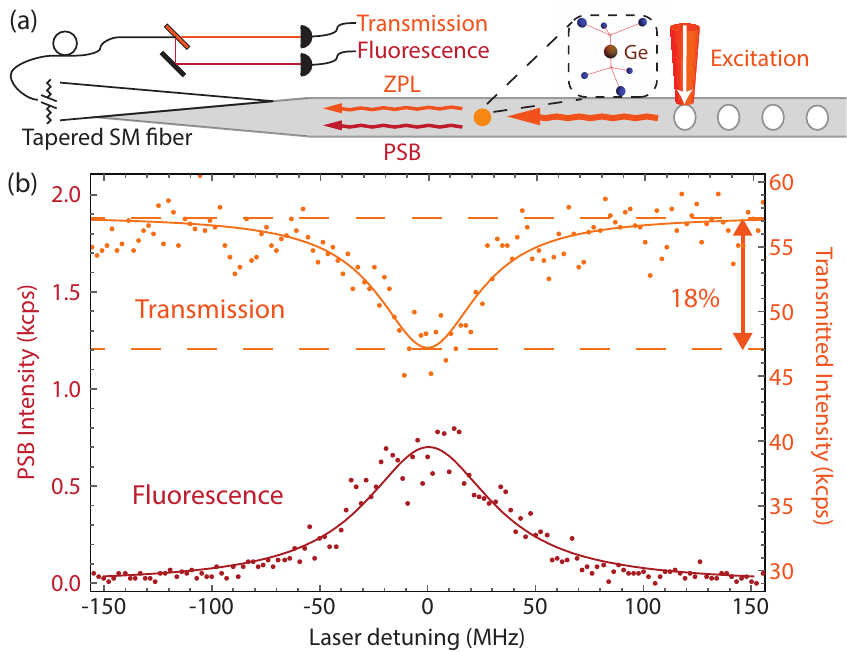}
		\caption{\label{fig4}
	(a) Schematic for single-pass transmission measurement. The resonant excitation ($I/I_{sat} \sim 0.02$) is focused on the Bragg mirror to scatter light into the waveguide. We collect transmitted light into the tapered optical fiber and subsequently separate the transmission and fluorescence (PSB) using a bandpass filter.
	(b) Transmission spectrum of the GeV-waveguide device, showing $18 \pm 1\%$ extinction on resonance. Transmission is shown on top (right axis) in orange and PSB fluorescence is shown on the bottom (left axis) in red. The solid curves are Lorentzian fits.
}
\end{figure}

\begin{figure}
	\includegraphics[width=\linewidth]{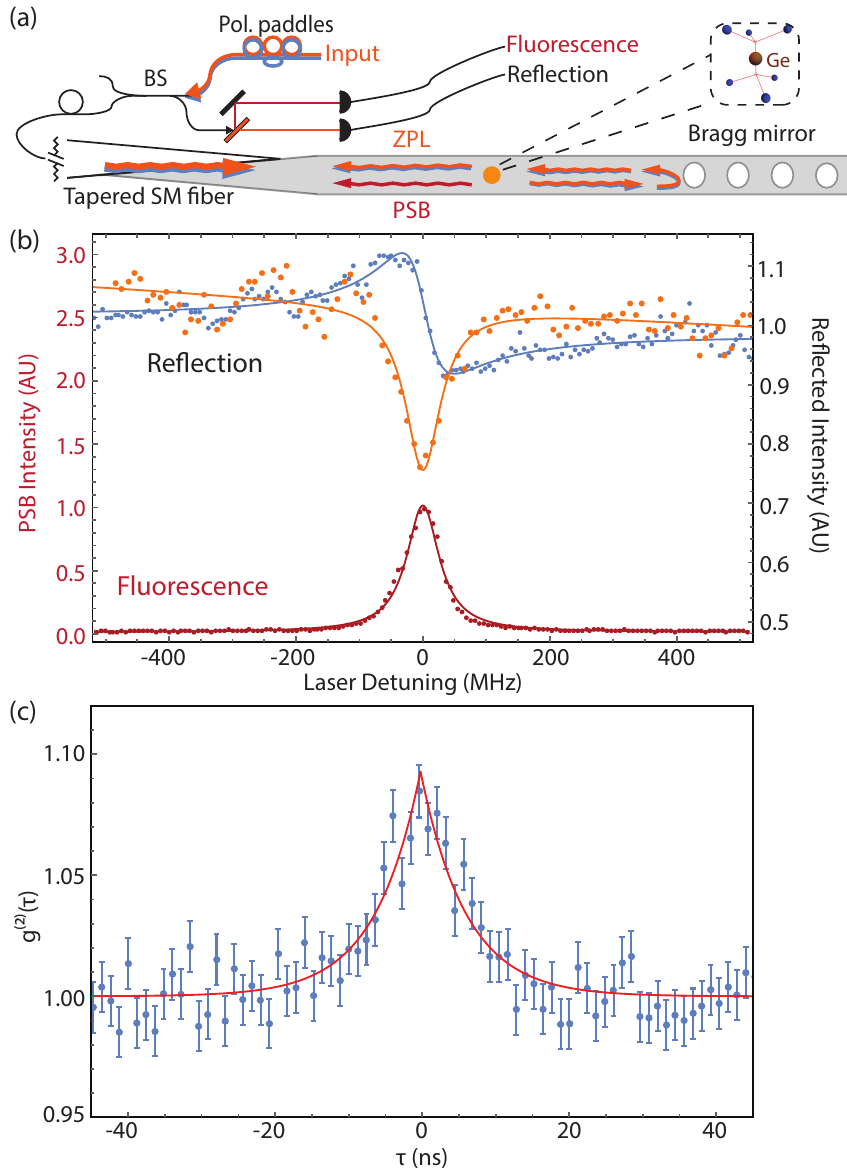}
		\caption{\label{fig5}
	(a) Schematic for homodyne interferometer. We excite ($I/I_{sat} \sim 0.02$) and collect through two ports of a fiber beamsplitter (BS). We use polarization paddles to change the excitation polarization.
	(b) Homodyne interfererometry with a single GeV. The reflected signal is shown on top (right axis) at two different input polarizations (orange and blue). PSB fluorescence is shown on the bottom (left axis) in red. The solid red curve is a Lorentzian fit and the solid blue and orange curves are fits to a phenomenological model \cite{SOM}.
	(c) Autocorrelation measurement. We perform Hanbury Brown-Twiss interferometry on the reflected field in the case of destructive interference for single photons. We measure bunching of $g^{(2)}(0) = 1.09 \pm 0.03$ with a decay time $\tau_b = 6.2 \pm 2.7$ \si{\ns}. The solid curve is an exponential fit \cite{SOM}.
	}
\end{figure}

In Fig. \ref{fig3}(b) we demonstrate coherent control over the GeV optical transition 1-3
by applying a resonant 40-ns pulse and observing optical Rabi oscillations using photons detected on the PSB. 
At high excitation power, the GeV optical transition undergoes spectral diffusion of roughly \SI{300}{\MHz} about the original resonance frequency.
In order to mitigate spectral diffusion at high excitation intensities, we use an active feedback sequence \cite{bernien2013heralded,goldman2015phonon} that stabilizes the GeV resonance frequency while maintaining a high duty cycle on resonance \cite{SOM}. 
This procedure enables high contrast oscillations at a Rabi frequency of $310 \pm 2$ \si{\MHz} with a decay time of $6.59 \pm 0.02$ \si{\ns} at \SI{5}{\K}, close to the excited state lifetime of $6.1 \pm 0.2$ \si{\ns}.
The decay rate of Rabi oscillations increases linearly with temperature [inset in Fig. \ref{fig3}(b)], suggesting that a single-phonon orbital relaxation process limits optical coherence at low temperatures between $T = $ \SI{5}{\K} and $T = $ \SI{10}{\K}, again similar to the case of the SiV \cite{jahnke2015electron}. 

These excellent optical properties allow us to observe the extinction of resonant transmission through a single
GeV center in a waveguide, as demonstrated in Fig. \ref{fig4}. We focus the excitation on the Bragg mirror in order to scatter laser light into the waveguide. 
We collect the light transmitted through the GeV into the tapered fiber, separating the transmitted and fluorescence (PSB) components using a bandpass filter [Fig. \ref{fig4}(a)]. We find that 
on resonance, a single GeV reduces waveguide transmission by $18 \pm 1 \%$ [Fig. \ref{fig4}(b)].
The extinction of resonant light by a single quantum emitter is an effective measure of the strength of emitter-photon interactions,
and is related to 
the emitter-waveguide cooperativity $C = \Gamma_{1D}/\Gamma'$, the ratio of the decay rate into the waveguide $\Gamma_{1D}$, to the sum of atomic decay rates to all other channels and dephasing $\Gamma'$ \cite{chang2007single,goban2015superradiance}.
From the measured extinction from a single GeV, we directly obtain the GeV-waveguide cooperativity of $C \geq 0.10\pm 0.01$ \cite{SOM}.
Because the GeV is a multi-level system with finite thermal population in level 2 at $T =$ \SI{5}{\K}, the cooperativity extracted from transmission of light resonant with transition 1-3 is a lower bound and can be improved by initializing the GeV in state 1 by optical pumping \cite{sipahigil2016integrated}.
The extinction of resonant light results from destructive interference between the driving field (a local oscillator) and resonance fluorescence from the GeV, and in general depends on the relative phase between them \cite{schulte2015quadrature}.
This phase can be controlled in a homodyne measurement involving a laser field and GeV resonance fluorescence
in the stable nanophotonic interferometer depicted in Fig. \ref{fig5}(a). Here, the laser light is injected 
through one port of a fiber beamsplitter connected to the tapered fiber and collected via a second beamsplitter port.
The reflected field at ZPL wavelengths consists of interference between GeV resonance fluorescence and the near-resonant excitation laser light reflected back into the fiber by the  Bragg mirror, which acts as a local oscillator.
We vary the relative amplitude and phase of the local oscillator with respect to the GeV resonance fluorescence by modifying the polarization of input laser light (for details, see \cite{SOM}).
Using this technique, we observe the change in lineshape of the output light from symmetric, corresponding to destructive interference (orange), to dispersive (blue) [Fig. \ref{fig5}(b)].

Finally, Fig. \ref{fig5}(c) demonstrates the quantum nonlinear character of the coupled GeV-waveguide system. 
In the homodyne measurement, the local oscillator is a weak coherent state with non-negligible single and two-photon components, whereas the GeV resonance fluorescence consists of only single photons.
In the case of large single atom-photon interaction probability, a single photon in the waveguide mode can saturate the GeV center and a single GeV can alter the photon statistics of the output field \cite{chang2007single}.
As an example, in 
the case of destructive interference between the two fields, the output field consists preferentially of two-photon components from the local oscillator, resulting in photon bunching. 
We probe the photon statistics of the homodyne output field using Hanbury Brown-Twiss interferometry and 
 observe, in the case of destructive interference, $g^{(2)}(0) = 1.09 \pm 0.03$ [Fig. \ref{fig5}(c)].
This observation provides direct evidence of device nonlinearity at the level of a single photon \cite{sipahigil2016integrated, javadi2015single, chang2007single}.

We next turn to the discussion of our experimental observations.
The GeV excited state lifetime [Fig. \ref{fig2}(d)] sets a theoretical upper bound on the single-photon flux from a GeV center of roughly $160$ Mcps.
From the saturation curve in Fig. \ref{fig2}(b), we infer that the maximum possible ZPL single-photon detection rate is $0.79 \pm 0.02$ Mcps in our experiment.
Accounting for the ZPL branching ratio and setup inefficiencies, we estimate that per excitation, the probability of emission of a photon into the waveguide mode is at least $0.1$ \cite{SOM, patel2016efficient}.
These measurements demonstrate that a single GeV center in a diamond waveguide is an efficient source of narrowband single photons.
The cooperativity measured in the transmission experiment (Fig. \ref{fig4}) is reduced by a combination of line-broadening mechanisms and multi-level dynamics \cite{sipahigil2016integrated, chang2007single}.
Since the branching ratios of the GeV optical transitions are not yet known, it is difficult to develop a comprehensive model of the population dynamics.
Using a simple three-level model, we estimate the phonon relaxation rate using $\gamma_{p}=2\gamma_{Rabi}-\frac{3}{2}\gamma_0$ 
\cite{goldman2015phonon}, where  $\gamma_{Rabi}$ is the decay rate of Rabi oscillations, and $\gamma_0$ ($\gamma_{p}$) is the excited state (phonon) relaxation rate. 
Using the measured value of $\gamma_{Rabi}/(2 \pi) = 24 \pm 0.1$ \si{\MHz} from Fig. \ref{fig3}(b), we infer that phonon relaxation leads to $\gamma_p/(2\pi) = 9 \pm 2$ \si{\MHz} of Markovian line-broadening at $T =$ \SI{5}{\K}.
Therefore, the observed \SI{73}{\MHz} linewidth is limited at \SI{5}{\K} by a combination of phonon relaxation and residual spectral diffusion and can likely be reduced further at lower temperatures, as demonstrated for GeV centers in bulk diamond \cite{siyushev2016optical}.

The present observations, together with recent advances involving SiV centers \cite{sipahigil2016integrated}, demonstrate significant potential for the realization of quantum nanophotonic devices using the family of color centers in diamond with inversion symmetry \cite{goss2005vacancy}.
The negatively-charged GeV center investigated here also has an electronic spin ($S = 1/2$) degree of freedom that can be manipulated using optical and microwave fields,
making it a promising spin-photon interface \cite{siyushev2016optical}.
As in the case of the SiV, coherence between GeV orbital and spin sublevels is limited by phonon relaxation at finite temperatures.
Ongoing efforts to suppress these relaxation processes at lower temperatures should result in long spin coherence times \cite{jahnke2015electron}.
Our observations also point to  some key differences in the optical properties of GeV and SiV centers. In particular,   the GeV excited state lifetime [Fig. \ref{fig2}(d)] shows negligible temperature dependence and high sensitivity to changes in the local photonic density of states, indicating the primarily radiative nature of the decay.  By contrast, the excited state lifetime of the SiV has strong temperature dependence \cite{jahnke2015electron} and does not respond as sensitively to changes in the local photonic density of states \cite{evans2016narrow}. These observations demonstrate that the GeV has a higher quantum efficiency than the SiV
and directly result in
the strong extinction of light in a single-pass, without the need for a slow-light waveguide or cavity.
%
%
In particular, the large extinction observed from a single GeV is competitive with single-pass transmission experiments using trapped atoms \cite{tey2008strong}, ions \cite{hetet2011single}, molecules \cite{wrigge2008efficient}, and quantum dots \cite{vamivakas2007strong}.

%
These observations open up exciting prospects for the realization of coherent quantum optical nodes with exceptionally strong atom-light coupling.
In particular, the high quantum efficiency of the GeV, when integrated into diamond nanocavities with previously demonstrated quality factor-mode volume ratios $Q/V > 10^4$ \cite{burek2014high, burek2016fiber}, could enable device cooperativities $C>100$, leading to deterministic single atom-photon interactions.
GeV orbital and spin coherence properties can be improved by cooling devices below \SI{1}{\K}, potentially yielding long-lived quantum memories \cite{jahnke2015electron}. 
Arrays of such strongly-coupled GeV-nanophotonic devices could be used as a basis for the realization of integrated quantum optical networks with applications in quantum information science \cite{kimble2008quantum} and studies of many-body physics with strongly interacting photons \cite{carusotto2013quantum}.
We thank D. Twitchen and M. Markham from Element Six Inc. for substrates and C. Meuwly and M. Goldman for experimental help. Financial support was provided by the NSF, the CUA, the AFOSR MURI, the ONR MURI, the DARPA QuINESS program, the ARL, and the Vannevar Bush Faculty Fellowship program. F.J. is affiliated with the Center for Integrated Quantum Science and Technology (IQst) in BadenW\"{u}rttemberg, Germany. Devices were fabricated at the Harvard CNS supported under NSF award ECS-0335765.

\bibliography{GeVbib}

\begin{thebibliography}{35}%
\makeatletter
\providecommand \@ifxundefined [1]{%
 \@ifx{#1\undefined}
}%
\providecommand \@ifnum [1]{%
 \ifnum #1\expandafter \@firstoftwo
 \else \expandafter \@secondoftwo
 \fi
}%
\providecommand \@ifx [1]{%
 \ifx #1\expandafter \@firstoftwo
 \else \expandafter \@secondoftwo
 \fi
}%
\providecommand \natexlab [1]{#1}%
\providecommand \enquote  [1]{``#1''}%
\providecommand \bibnamefont  [1]{#1}%
\providecommand \bibfnamefont [1]{#1}%
\providecommand \citenamefont [1]{#1}%
\providecommand \href@noop [0]{\@secondoftwo}%
\providecommand \href [0]{\begingroup \@sanitize@url \@href}%
\providecommand \@href[1]{\@@startlink{#1}\@@href}%
\providecommand \@@href[1]{\endgroup#1\@@endlink}%
\providecommand \@sanitize@url [0]{\catcode `\\12\catcode `\$12\catcode
  `\&12\catcode `\#12\catcode `\^12\catcode `\_12\catcode `\%12\relax}%
\providecommand \@@startlink[1]{}%
\providecommand \@@endlink[0]{}%
\providecommand \url  [0]{\begingroup\@sanitize@url \@url }%
\providecommand \@url [1]{\endgroup\@href {#1}{\urlprefix }}%
\providecommand \urlprefix  [0]{URL }%
\providecommand \Eprint [0]{\href }%
\providecommand \doibase [0]{http://dx.doi.org/}%
\providecommand \selectlanguage [0]{\@gobble}%
\providecommand \bibinfo  [0]{\@secondoftwo}%
\providecommand \bibfield  [0]{\@secondoftwo}%
\providecommand \translation [1]{[#1]}%
\providecommand \BibitemOpen [0]{}%
\providecommand \bibitemStop [0]{}%
\providecommand \bibitemNoStop [0]{.\EOS\space}%
\providecommand \EOS [0]{\spacefactor3000\relax}%
\providecommand \BibitemShut  [1]{\csname bibitem#1\endcsname}%
\let\auto@bib@innerbib\@empty
\bibitem [{\citenamefont {Kimble}(2008)}]{kimble2008quantum}%
  \BibitemOpen
  \bibfield  {author} {\bibinfo {author} {\bibfnamefont {H.~J.}\ \bibnamefont
  {Kimble}},\ }\href@noop {} {\bibfield  {journal} {\bibinfo  {journal}
  {Nature}\ }\textbf {\bibinfo {volume} {453}},\ \bibinfo {pages} {1023}
  (\bibinfo {year} {2008})}\BibitemShut {NoStop}%
\bibitem [{\citenamefont {Lodahl}\ \emph {et~al.}(2015)\citenamefont {Lodahl},
  \citenamefont {Mahmoodian},\ and\ \citenamefont
  {Stobbe}}]{lodahl2015interfacing}%
  \BibitemOpen
  \bibfield  {author} {\bibinfo {author} {\bibfnamefont {P.}~\bibnamefont
  {Lodahl}}, \bibinfo {author} {\bibfnamefont {S.}~\bibnamefont {Mahmoodian}},
  \ and\ \bibinfo {author} {\bibfnamefont {S.}~\bibnamefont {Stobbe}},\
  }\href@noop {} {\bibfield  {journal} {\bibinfo  {journal} {Rev. Mod. Phys.}\
  }\textbf {\bibinfo {volume} {87}},\ \bibinfo {pages} {347} (\bibinfo {year}
  {2015})}\BibitemShut {NoStop}%
\bibitem [{\citenamefont {Carusotto}\ and\ \citenamefont
  {Ciuti}(2013)}]{carusotto2013quantum}%
  \BibitemOpen
  \bibfield  {author} {\bibinfo {author} {\bibfnamefont {I.}~\bibnamefont
  {Carusotto}}\ and\ \bibinfo {author} {\bibfnamefont {C.}~\bibnamefont
  {Ciuti}},\ }\href@noop {} {\bibfield  {journal} {\bibinfo  {journal} {Rev.
  Mod. Phys.}\ }\textbf {\bibinfo {volume} {85}},\ \bibinfo {pages} {299}
  (\bibinfo {year} {2013})}\BibitemShut {NoStop}%
\bibitem [{\citenamefont {Aharonovich}\ \emph {et~al.}(2016)\citenamefont
  {Aharonovich}, \citenamefont {Englund},\ and\ \citenamefont
  {Toth}}]{aharonovich2016solid}%
  \BibitemOpen
  \bibfield  {author} {\bibinfo {author} {\bibfnamefont {I.}~\bibnamefont
  {Aharonovich}}, \bibinfo {author} {\bibfnamefont {D.}~\bibnamefont
  {Englund}}, \ and\ \bibinfo {author} {\bibfnamefont {M.}~\bibnamefont
  {Toth}},\ }\href@noop {} {\bibfield  {journal} {\bibinfo  {journal} {Nat.
  Photon.}\ }\textbf {\bibinfo {volume} {10}},\ \bibinfo {pages} {631}
  (\bibinfo {year} {2016})}\BibitemShut {NoStop}%
\bibitem [{\citenamefont {Maurer}\ \emph {et~al.}(2012)\citenamefont {Maurer},
  \citenamefont {Kucsko}, \citenamefont {Latta}, \citenamefont {Jiang},
  \citenamefont {Yao}, \citenamefont {Bennett}, \citenamefont {Pastawski},
  \citenamefont {Hunger}, \citenamefont {Chisholm}, \citenamefont {Markham}
  \emph {et~al.}}]{maurer2012room}%
  \BibitemOpen
  \bibfield  {author} {\bibinfo {author} {\bibfnamefont {P.~C.}\ \bibnamefont
  {Maurer}}, \bibinfo {author} {\bibfnamefont {G.}~\bibnamefont {Kucsko}},
  \bibinfo {author} {\bibfnamefont {C.}~\bibnamefont {Latta}}, \bibinfo
  {author} {\bibfnamefont {L.}~\bibnamefont {Jiang}}, \bibinfo {author}
  {\bibfnamefont {N.~Y.}\ \bibnamefont {Yao}}, \bibinfo {author} {\bibfnamefont
  {S.~D.}\ \bibnamefont {Bennett}}, \bibinfo {author} {\bibfnamefont
  {F.}~\bibnamefont {Pastawski}}, \bibinfo {author} {\bibfnamefont
  {D.}~\bibnamefont {Hunger}}, \bibinfo {author} {\bibfnamefont
  {N.}~\bibnamefont {Chisholm}}, \bibinfo {author} {\bibfnamefont
  {M.}~\bibnamefont {Markham}},  \emph {et~al.},\ }\href@noop {} {\bibfield
  {journal} {\bibinfo  {journal} {Science}\ }\textbf {\bibinfo {volume}
  {336}},\ \bibinfo {pages} {1283} (\bibinfo {year} {2012})}\BibitemShut
  {NoStop}%
\bibitem [{\citenamefont {Evans}\ \emph {et~al.}(2016)\citenamefont {Evans},
  \citenamefont {Sipahigil}, \citenamefont {Sukachev}, \citenamefont {Zibrov},\
  and\ \citenamefont {Lukin}}]{evans2016narrow}%
  \BibitemOpen
  \bibfield  {author} {\bibinfo {author} {\bibfnamefont {R.~E.}\ \bibnamefont
  {Evans}}, \bibinfo {author} {\bibfnamefont {A.}~\bibnamefont {Sipahigil}},
  \bibinfo {author} {\bibfnamefont {D.~D.}\ \bibnamefont {Sukachev}}, \bibinfo
  {author} {\bibfnamefont {A.~S.}\ \bibnamefont {Zibrov}}, \ and\ \bibinfo
  {author} {\bibfnamefont {M.~D.}\ \bibnamefont {Lukin}},\ }\href@noop {}
  {\bibfield  {journal} {\bibinfo  {journal} {Phys. Rev. Appl.}\ }\textbf
  {\bibinfo {volume} {5}},\ \bibinfo {pages} {044010} (\bibinfo {year}
  {2016})}\BibitemShut {NoStop}%
\bibitem [{\citenamefont {Sipahigil}\ \emph {et~al.}(2016)\citenamefont
  {Sipahigil}, \citenamefont {Evans}, \citenamefont {Sukachev}, \citenamefont
  {Burek}, \citenamefont {Borregaard}, \citenamefont {Bhaskar}, \citenamefont
  {Nguyen}, \citenamefont {Pacheco}, \citenamefont {Atikian}, \citenamefont
  {Meuwly} \emph {et~al.}}]{sipahigil2016integrated}%
  \BibitemOpen
  \bibfield  {author} {\bibinfo {author} {\bibfnamefont {A.}~\bibnamefont
  {Sipahigil}}, \bibinfo {author} {\bibfnamefont {R.~E.}\ \bibnamefont
  {Evans}}, \bibinfo {author} {\bibfnamefont {D.~D.}\ \bibnamefont {Sukachev}},
  \bibinfo {author} {\bibfnamefont {M.~J.}\ \bibnamefont {Burek}}, \bibinfo
  {author} {\bibfnamefont {J.}~\bibnamefont {Borregaard}}, \bibinfo {author}
  {\bibfnamefont {M.~K.}\ \bibnamefont {Bhaskar}}, \bibinfo {author}
  {\bibfnamefont {C.~T.}\ \bibnamefont {Nguyen}}, \bibinfo {author}
  {\bibfnamefont {J.~L.}\ \bibnamefont {Pacheco}}, \bibinfo {author}
  {\bibfnamefont {H.~A.}\ \bibnamefont {Atikian}}, \bibinfo {author}
  {\bibfnamefont {C.}~\bibnamefont {Meuwly}},  \emph {et~al.},\ }\href
  {\doibase 10.1126/science.aah6875} {\bibfield  {journal} {\bibinfo  {journal}
  {Science}\ }\textbf {\bibinfo {volume} {354}},\ \bibinfo {pages} {847}
  (\bibinfo {year} {2016})}\BibitemShut {NoStop}%
\bibitem [{\citenamefont {Jantzen}\ \emph {et~al.}(2016)\citenamefont
  {Jantzen}, \citenamefont {Kurz}, \citenamefont {Rudnicki}, \citenamefont
  {Sch{\"a}fermeier}, \citenamefont {Jahnke}, \citenamefont {Andersen},
  \citenamefont {Davydov}, \citenamefont {Agafonov}, \citenamefont {Kubanek},
  \citenamefont {Rogers},\ and\ \citenamefont
  {Jelezko}}]{jantzen2016nanodiamonds}%
  \BibitemOpen
  \bibfield  {author} {\bibinfo {author} {\bibfnamefont {U.}~\bibnamefont
  {Jantzen}}, \bibinfo {author} {\bibfnamefont {A.~B.}\ \bibnamefont {Kurz}},
  \bibinfo {author} {\bibfnamefont {D.~S.}\ \bibnamefont {Rudnicki}}, \bibinfo
  {author} {\bibfnamefont {C.}~\bibnamefont {Sch{\"a}fermeier}}, \bibinfo
  {author} {\bibfnamefont {K.~D.}\ \bibnamefont {Jahnke}}, \bibinfo {author}
  {\bibfnamefont {U.~L.}\ \bibnamefont {Andersen}}, \bibinfo {author}
  {\bibfnamefont {V.~A.}\ \bibnamefont {Davydov}}, \bibinfo {author}
  {\bibfnamefont {V.~N.}\ \bibnamefont {Agafonov}}, \bibinfo {author}
  {\bibfnamefont {A.}~\bibnamefont {Kubanek}}, \bibinfo {author} {\bibfnamefont
  {L.~J.}\ \bibnamefont {Rogers}}, \ and\ \bibinfo {author} {\bibfnamefont
  {F.}~\bibnamefont {Jelezko}},\ }\href
  {http://stacks.iop.org/1367-2630/18/i=7/a=073036} {\bibfield  {journal}
  {\bibinfo  {journal} {New J. Phys.}\ }\textbf {\bibinfo {volume} {18}},\
  \bibinfo {pages} {073036} (\bibinfo {year} {2016})}\BibitemShut {NoStop}%
\bibitem [{\citenamefont {Li}\ \emph {et~al.}(2016)\citenamefont {Li},
  \citenamefont {Zhou}, \citenamefont {Rasmita}, \citenamefont {Aharonovich},\
  and\ \citenamefont {Gao}}]{li2016nonblinking}%
  \BibitemOpen
  \bibfield  {author} {\bibinfo {author} {\bibfnamefont {K.}~\bibnamefont
  {Li}}, \bibinfo {author} {\bibfnamefont {Y.}~\bibnamefont {Zhou}}, \bibinfo
  {author} {\bibfnamefont {A.}~\bibnamefont {Rasmita}}, \bibinfo {author}
  {\bibfnamefont {I.}~\bibnamefont {Aharonovich}}, \ and\ \bibinfo {author}
  {\bibfnamefont {W.~B.}\ \bibnamefont {Gao}},\ }\href@noop {} {\bibfield
  {journal} {\bibinfo  {journal} {Phys. Rev. Appl.}\ }\textbf {\bibinfo
  {volume} {6}},\ \bibinfo {pages} {024010} (\bibinfo {year}
  {2016})}\BibitemShut {NoStop}%
\bibitem [{\citenamefont {Hepp}\ \emph {et~al.}(2014)\citenamefont {Hepp},
  \citenamefont {M\"uller}, \citenamefont {Waselowski}, \citenamefont {Becker},
  \citenamefont {Pingault}, \citenamefont {Sternschulte}, \citenamefont
  {Steinm\"uller-Nethl}, \citenamefont {Gali}, \citenamefont {Maze},
  \citenamefont {Atat\"ure},\ and\ \citenamefont
  {Becher}}]{hepp2014electronic}%
  \BibitemOpen
  \bibfield  {author} {\bibinfo {author} {\bibfnamefont {C.}~\bibnamefont
  {Hepp}}, \bibinfo {author} {\bibfnamefont {T.}~\bibnamefont {M\"uller}},
  \bibinfo {author} {\bibfnamefont {V.}~\bibnamefont {Waselowski}}, \bibinfo
  {author} {\bibfnamefont {J.~N.}\ \bibnamefont {Becker}}, \bibinfo {author}
  {\bibfnamefont {B.}~\bibnamefont {Pingault}}, \bibinfo {author}
  {\bibfnamefont {H.}~\bibnamefont {Sternschulte}}, \bibinfo {author}
  {\bibfnamefont {D.}~\bibnamefont {Steinm\"uller-Nethl}}, \bibinfo {author}
  {\bibfnamefont {A.}~\bibnamefont {Gali}}, \bibinfo {author} {\bibfnamefont
  {J.~R.}\ \bibnamefont {Maze}}, \bibinfo {author} {\bibfnamefont
  {M.}~\bibnamefont {Atat\"ure}}, \ and\ \bibinfo {author} {\bibfnamefont
  {C.}~\bibnamefont {Becher}},\ }\href {\doibase
  10.1103/PhysRevLett.112.036405} {\bibfield  {journal} {\bibinfo  {journal}
  {Phys. Rev. Lett.}\ }\textbf {\bibinfo {volume} {112}},\ \bibinfo {pages}
  {036405} (\bibinfo {year} {2014})}\BibitemShut {NoStop}%
\bibitem [{\citenamefont {Goss}\ \emph {et~al.}(2005)\citenamefont {Goss},
  \citenamefont {Briddon}, \citenamefont {Rayson}, \citenamefont {Sque},\ and\
  \citenamefont {Jones}}]{goss2005vacancy}%
  \BibitemOpen
  \bibfield  {author} {\bibinfo {author} {\bibfnamefont {J.~P.}\ \bibnamefont
  {Goss}}, \bibinfo {author} {\bibfnamefont {P.~R.}\ \bibnamefont {Briddon}},
  \bibinfo {author} {\bibfnamefont {M.~J.}\ \bibnamefont {Rayson}}, \bibinfo
  {author} {\bibfnamefont {S.~J.}\ \bibnamefont {Sque}}, \ and\ \bibinfo
  {author} {\bibfnamefont {R.}~\bibnamefont {Jones}},\ }\href@noop {}
  {\bibfield  {journal} {\bibinfo  {journal} {Phys. Rev. B}\ }\textbf {\bibinfo
  {volume} {72}},\ \bibinfo {pages} {035214} (\bibinfo {year}
  {2005})}\BibitemShut {NoStop}%
\bibitem [{\citenamefont {Palyanov}\ \emph {et~al.}(2015)\citenamefont
  {Palyanov}, \citenamefont {Kupriyanov}, \citenamefont {Borzdov},\ and\
  \citenamefont {Surovtsev}}]{palyanov2015germanium}%
  \BibitemOpen
  \bibfield  {author} {\bibinfo {author} {\bibfnamefont {Y.~N.}\ \bibnamefont
  {Palyanov}}, \bibinfo {author} {\bibfnamefont {I.~N.}\ \bibnamefont
  {Kupriyanov}}, \bibinfo {author} {\bibfnamefont {Y.~M.}\ \bibnamefont
  {Borzdov}}, \ and\ \bibinfo {author} {\bibfnamefont {N.~V.}\ \bibnamefont
  {Surovtsev}},\ }\href@noop {} {\bibfield  {journal} {\bibinfo  {journal}
  {Sci. Rep.}\ }\textbf {\bibinfo {volume} {5}},\ \bibinfo {pages} {14789}
  (\bibinfo {year} {2015})}\BibitemShut {NoStop}%
\bibitem [{\citenamefont {Iwasaki}\ \emph {et~al.}(2015)\citenamefont
  {Iwasaki}, \citenamefont {Ishibashi}, \citenamefont {Miyamoto}, \citenamefont
  {Doi}, \citenamefont {Kobayashi}, \citenamefont {Miyazaki}, \citenamefont
  {Tahara}, \citenamefont {Jahnke}, \citenamefont {Rogers}, \citenamefont
  {Naydenov} \emph {et~al.}}]{iwasaki2015germanium}%
  \BibitemOpen
  \bibfield  {author} {\bibinfo {author} {\bibfnamefont {T.}~\bibnamefont
  {Iwasaki}}, \bibinfo {author} {\bibfnamefont {F.}~\bibnamefont {Ishibashi}},
  \bibinfo {author} {\bibfnamefont {Y.}~\bibnamefont {Miyamoto}}, \bibinfo
  {author} {\bibfnamefont {Y.}~\bibnamefont {Doi}}, \bibinfo {author}
  {\bibfnamefont {S.}~\bibnamefont {Kobayashi}}, \bibinfo {author}
  {\bibfnamefont {T.}~\bibnamefont {Miyazaki}}, \bibinfo {author}
  {\bibfnamefont {K.}~\bibnamefont {Tahara}}, \bibinfo {author} {\bibfnamefont
  {K.~D.}\ \bibnamefont {Jahnke}}, \bibinfo {author} {\bibfnamefont {L.~J.}\
  \bibnamefont {Rogers}}, \bibinfo {author} {\bibfnamefont {B.}~\bibnamefont
  {Naydenov}},  \emph {et~al.},\ }\href@noop {} {\bibfield  {journal} {\bibinfo
   {journal} {Sci. Rep.}\ }\textbf {\bibinfo {volume} {5}},\ \bibinfo {pages}
  {12882} (\bibinfo {year} {2015})}\BibitemShut {NoStop}%
\bibitem [{\citenamefont {Ekimov}\ \emph {et~al.}(2015)\citenamefont {Ekimov},
  \citenamefont {Lyapin}, \citenamefont {Boldyrev}, \citenamefont {Kondrin},
  \citenamefont {Khmelnitskiy}, \citenamefont {Gavva}, \citenamefont
  {Kotereva},\ and\ \citenamefont {Popova}}]{ekimov2015germanium}%
  \BibitemOpen
  \bibfield  {author} {\bibinfo {author} {\bibfnamefont {E.~A.}\ \bibnamefont
  {Ekimov}}, \bibinfo {author} {\bibfnamefont {S.}~\bibnamefont {Lyapin}},
  \bibinfo {author} {\bibfnamefont {K.~N.}\ \bibnamefont {Boldyrev}}, \bibinfo
  {author} {\bibfnamefont {M.~V.}\ \bibnamefont {Kondrin}}, \bibinfo {author}
  {\bibfnamefont {R.}~\bibnamefont {Khmelnitskiy}}, \bibinfo {author}
  {\bibfnamefont {V.~A.}\ \bibnamefont {Gavva}}, \bibinfo {author}
  {\bibfnamefont {T.~V.}\ \bibnamefont {Kotereva}}, \ and\ \bibinfo {author}
  {\bibfnamefont {M.~N.}\ \bibnamefont {Popova}},\ }\href@noop {} {\bibfield
  {journal} {\bibinfo  {journal} {JETP Lett.}\ }\textbf {\bibinfo {volume}
  {102}},\ \bibinfo {pages} {701} (\bibinfo {year} {2015})}\BibitemShut
  {NoStop}%
\bibitem [{\citenamefont {Palyanov}\ \emph {et~al.}(2016)\citenamefont
  {Palyanov}, \citenamefont {Kupriyanov}, \citenamefont {Borzdov},
  \citenamefont {Khokhryakov},\ and\ \citenamefont
  {Surovtsev}}]{palyanov2016high}%
  \BibitemOpen
  \bibfield  {author} {\bibinfo {author} {\bibfnamefont {Y.~N.}\ \bibnamefont
  {Palyanov}}, \bibinfo {author} {\bibfnamefont {I.~N.}\ \bibnamefont
  {Kupriyanov}}, \bibinfo {author} {\bibfnamefont {Y.~M.}\ \bibnamefont
  {Borzdov}}, \bibinfo {author} {\bibfnamefont {A.~F.}\ \bibnamefont
  {Khokhryakov}}, \ and\ \bibinfo {author} {\bibfnamefont {N.~V.}\ \bibnamefont
  {Surovtsev}},\ }\href@noop {} {\bibfield  {journal} {\bibinfo  {journal}
  {Cryst. Growth Des.}\ }\textbf {\bibinfo {volume} {16}},\ \bibinfo {pages}
  {3510} (\bibinfo {year} {2016})}\BibitemShut {NoStop}%
\bibitem [{\citenamefont {M{\"u}ller}\ \emph {et~al.}(2014)\citenamefont
  {M{\"u}ller}, \citenamefont {Hepp}, \citenamefont {Pingault}, \citenamefont
  {Neu}, \citenamefont {Gsell}, \citenamefont {Schreck}, \citenamefont
  {Sternschulte}, \citenamefont {Steinm{\"u}ller-Nethl}, \citenamefont
  {Becher},\ and\ \citenamefont {Atat{\"u}re}}]{muller2014optical}%
  \BibitemOpen
  \bibfield  {author} {\bibinfo {author} {\bibfnamefont {T.}~\bibnamefont
  {M{\"u}ller}}, \bibinfo {author} {\bibfnamefont {C.}~\bibnamefont {Hepp}},
  \bibinfo {author} {\bibfnamefont {B.}~\bibnamefont {Pingault}}, \bibinfo
  {author} {\bibfnamefont {E.}~\bibnamefont {Neu}}, \bibinfo {author}
  {\bibfnamefont {S.}~\bibnamefont {Gsell}}, \bibinfo {author} {\bibfnamefont
  {M.}~\bibnamefont {Schreck}}, \bibinfo {author} {\bibfnamefont
  {H.}~\bibnamefont {Sternschulte}}, \bibinfo {author} {\bibfnamefont
  {D.}~\bibnamefont {Steinm{\"u}ller-Nethl}}, \bibinfo {author} {\bibfnamefont
  {C.}~\bibnamefont {Becher}}, \ and\ \bibinfo {author} {\bibfnamefont
  {M.}~\bibnamefont {Atat{\"u}re}},\ }\href@noop {} {\bibfield  {journal}
  {\bibinfo  {journal} {Nat. Comm.}\ }\textbf {\bibinfo {volume} {5}},\
  \bibinfo {pages} {3328} (\bibinfo {year} {2014})}\BibitemShut {NoStop}%
\bibitem [{\citenamefont {Siyushev}\ \emph {et~al.}(2016)\citenamefont
  {Siyushev}, \citenamefont {Metsch}, \citenamefont {Ijaz}, \citenamefont
  {Binder}, \citenamefont {Bhaskar}, \citenamefont {Sukachev}, \citenamefont
  {Sipahigil}, \citenamefont {Evans}, \citenamefont {Nguyen}, \citenamefont
  {Lukin} \emph {et~al.}}]{siyushev2016optical}%
  \BibitemOpen
  \bibfield  {author} {\bibinfo {author} {\bibfnamefont {P.}~\bibnamefont
  {Siyushev}}, \bibinfo {author} {\bibfnamefont {M.~H.}\ \bibnamefont
  {Metsch}}, \bibinfo {author} {\bibfnamefont {A.}~\bibnamefont {Ijaz}},
  \bibinfo {author} {\bibfnamefont {J.~M.}\ \bibnamefont {Binder}}, \bibinfo
  {author} {\bibfnamefont {M.~K.}\ \bibnamefont {Bhaskar}}, \bibinfo {author}
  {\bibfnamefont {D.~D.}\ \bibnamefont {Sukachev}}, \bibinfo {author}
  {\bibfnamefont {A.}~\bibnamefont {Sipahigil}}, \bibinfo {author}
  {\bibfnamefont {R.~E.}\ \bibnamefont {Evans}}, \bibinfo {author}
  {\bibfnamefont {C.~T.}\ \bibnamefont {Nguyen}}, \bibinfo {author}
  {\bibfnamefont {M.~D.}\ \bibnamefont {Lukin}},  \emph {et~al.},\ }\href@noop
  {} {\bibfield  {journal} {\bibinfo  {journal} {arXiv preprint
  arXiv:1612.02947}\ } (\bibinfo {year} {2016})}\BibitemShut {NoStop}%
\bibitem [{SOM()}]{SOM}%
  \BibitemOpen
  \href@noop {} {}\bibinfo {note} {See supplemental materials online for
  details about the experimental setup, the GeV-waveguide coupling efficiency,
  fits to data in the main text, calculation of cooperativity, and operating
  principles of homodyne measurement.}\BibitemShut {Stop}%
\bibitem [{\citenamefont {Chang}\ \emph {et~al.}(2007)\citenamefont {Chang},
  \citenamefont {S{\o}rensen}, \citenamefont {Demler},\ and\ \citenamefont
  {Lukin}}]{chang2007single}%
  \BibitemOpen
  \bibfield  {author} {\bibinfo {author} {\bibfnamefont {D.~E.}\ \bibnamefont
  {Chang}}, \bibinfo {author} {\bibfnamefont {A.~S.}\ \bibnamefont
  {S{\o}rensen}}, \bibinfo {author} {\bibfnamefont {E.~A.}\ \bibnamefont
  {Demler}}, \ and\ \bibinfo {author} {\bibfnamefont {M.~D.}\ \bibnamefont
  {Lukin}},\ }\href@noop {} {\bibfield  {journal} {\bibinfo  {journal} {Nat.
  Phys.}\ }\textbf {\bibinfo {volume} {3}},\ \bibinfo {pages} {807} (\bibinfo
  {year} {2007})}\BibitemShut {NoStop}%
\bibitem [{\citenamefont {Goban}\ \emph {et~al.}(2015)\citenamefont {Goban},
  \citenamefont {Hung}, \citenamefont {Hood}, \citenamefont {Yu}, \citenamefont
  {Muniz}, \citenamefont {Painter},\ and\ \citenamefont
  {Kimble}}]{goban2015superradiance}%
  \BibitemOpen
  \bibfield  {author} {\bibinfo {author} {\bibfnamefont {A.}~\bibnamefont
  {Goban}}, \bibinfo {author} {\bibfnamefont {C.~L.}\ \bibnamefont {Hung}},
  \bibinfo {author} {\bibfnamefont {J.~D.}\ \bibnamefont {Hood}}, \bibinfo
  {author} {\bibfnamefont {S.~P.}\ \bibnamefont {Yu}}, \bibinfo {author}
  {\bibfnamefont {J.~A.}\ \bibnamefont {Muniz}}, \bibinfo {author}
  {\bibfnamefont {O.}~\bibnamefont {Painter}}, \ and\ \bibinfo {author}
  {\bibfnamefont {H.~J.}\ \bibnamefont {Kimble}},\ }\href {\doibase
  10.1103/PhysRevLett.115.063601} {\bibfield  {journal} {\bibinfo  {journal}
  {Phys. Rev. Lett.}\ }\textbf {\bibinfo {volume} {115}},\ \bibinfo {pages}
  {063601} (\bibinfo {year} {2015})}\BibitemShut {NoStop}%
\bibitem [{\citenamefont {Javadi}\ \emph {et~al.}(2015)\citenamefont {Javadi},
  \citenamefont {S{\"o}llner}, \citenamefont {Arcari}, \citenamefont {Hansen},
  \citenamefont {Midolo}, \citenamefont {Mahmoodian}, \citenamefont
  {Kir{\v{s}}ansk{\.e}}, \citenamefont {Pregnolato}, \citenamefont {Lee},
  \citenamefont {Song} \emph {et~al.}}]{javadi2015single}%
  \BibitemOpen
  \bibfield  {author} {\bibinfo {author} {\bibfnamefont {A.}~\bibnamefont
  {Javadi}}, \bibinfo {author} {\bibfnamefont {I.}~\bibnamefont {S{\"o}llner}},
  \bibinfo {author} {\bibfnamefont {M.}~\bibnamefont {Arcari}}, \bibinfo
  {author} {\bibfnamefont {S.~L.}\ \bibnamefont {Hansen}}, \bibinfo {author}
  {\bibfnamefont {L.}~\bibnamefont {Midolo}}, \bibinfo {author} {\bibfnamefont
  {S.}~\bibnamefont {Mahmoodian}}, \bibinfo {author} {\bibfnamefont
  {G.}~\bibnamefont {Kir{\v{s}}ansk{\.e}}}, \bibinfo {author} {\bibfnamefont
  {T.}~\bibnamefont {Pregnolato}}, \bibinfo {author} {\bibfnamefont
  {E.}~\bibnamefont {Lee}}, \bibinfo {author} {\bibfnamefont {J.}~\bibnamefont
  {Song}},  \emph {et~al.},\ }\href@noop {} {\bibfield  {journal} {\bibinfo
  {journal} {Nat. Comm.}\ }\textbf {\bibinfo {volume} {6}},\ \bibinfo {pages}
  {8655} (\bibinfo {year} {2015})}\BibitemShut {NoStop}%
\bibitem [{\citenamefont {Burek}\ \emph {et~al.}(2014)\citenamefont {Burek},
  \citenamefont {Chu}, \citenamefont {Liddy}, \citenamefont {Patel},
  \citenamefont {Rochman}, \citenamefont {Meesala}, \citenamefont {Hong},
  \citenamefont {Quan}, \citenamefont {Lukin},\ and\ \citenamefont
  {Lon{\v{c}}ar}}]{burek2014high}%
  \BibitemOpen
  \bibfield  {author} {\bibinfo {author} {\bibfnamefont {M.~J.}\ \bibnamefont
  {Burek}}, \bibinfo {author} {\bibfnamefont {Y.}~\bibnamefont {Chu}}, \bibinfo
  {author} {\bibfnamefont {M.~S.}\ \bibnamefont {Liddy}}, \bibinfo {author}
  {\bibfnamefont {P.}~\bibnamefont {Patel}}, \bibinfo {author} {\bibfnamefont
  {J.}~\bibnamefont {Rochman}}, \bibinfo {author} {\bibfnamefont
  {S.}~\bibnamefont {Meesala}}, \bibinfo {author} {\bibfnamefont
  {W.}~\bibnamefont {Hong}}, \bibinfo {author} {\bibfnamefont {Q.}~\bibnamefont
  {Quan}}, \bibinfo {author} {\bibfnamefont {M.~D.}\ \bibnamefont {Lukin}}, \
  and\ \bibinfo {author} {\bibfnamefont {M.}~\bibnamefont {Lon{\v{c}}ar}},\
  }\href@noop {} {\bibfield  {journal} {\bibinfo  {journal} {Nat. Comm.}\
  }\textbf {\bibinfo {volume} {5}},\ \bibinfo {pages} {5718} (\bibinfo {year}
  {2014})}\BibitemShut {NoStop}%
\bibitem [{\citenamefont {Burek}\ \emph {et~al.}(2016)\citenamefont {Burek},
  \citenamefont {Meuwly}, \citenamefont {Evans}, \citenamefont {Bhaskar},
  \citenamefont {Sipahigil}, \citenamefont {Meesala}, \citenamefont {Sukachev},
  \citenamefont {Nguyen}, \citenamefont {Pacheco}, \citenamefont {Bielejec}
  \emph {et~al.}}]{burek2016fiber}%
  \BibitemOpen
  \bibfield  {author} {\bibinfo {author} {\bibfnamefont {M.~J.}\ \bibnamefont
  {Burek}}, \bibinfo {author} {\bibfnamefont {C.}~\bibnamefont {Meuwly}},
  \bibinfo {author} {\bibfnamefont {R.~E.}\ \bibnamefont {Evans}}, \bibinfo
  {author} {\bibfnamefont {M.~K.}\ \bibnamefont {Bhaskar}}, \bibinfo {author}
  {\bibfnamefont {A.}~\bibnamefont {Sipahigil}}, \bibinfo {author}
  {\bibfnamefont {S.}~\bibnamefont {Meesala}}, \bibinfo {author} {\bibfnamefont
  {D.~D.}\ \bibnamefont {Sukachev}}, \bibinfo {author} {\bibfnamefont {C.~T.}\
  \bibnamefont {Nguyen}}, \bibinfo {author} {\bibfnamefont {J.~L.}\
  \bibnamefont {Pacheco}}, \bibinfo {author} {\bibfnamefont {E.}~\bibnamefont
  {Bielejec}},  \emph {et~al.},\ }\href@noop {} {\bibfield  {journal} {\bibinfo
   {journal} {arXiv preprint arXiv:1612.05285}\ } (\bibinfo {year}
  {2016})}\BibitemShut {NoStop}%
\bibitem [{\citenamefont {Tiecke}\ \emph {et~al.}(2015)\citenamefont {Tiecke},
  \citenamefont {Nayak}, \citenamefont {Thompson}, \citenamefont {Peyronel},
  \citenamefont {de~Leon}, \citenamefont {Vuleti{\'c}},\ and\ \citenamefont
  {Lukin}}]{tiecke2015efficient}%
  \BibitemOpen
  \bibfield  {author} {\bibinfo {author} {\bibfnamefont {T.~G.}\ \bibnamefont
  {Tiecke}}, \bibinfo {author} {\bibfnamefont {K.~P.}\ \bibnamefont {Nayak}},
  \bibinfo {author} {\bibfnamefont {J.~D.}\ \bibnamefont {Thompson}}, \bibinfo
  {author} {\bibfnamefont {T.}~\bibnamefont {Peyronel}}, \bibinfo {author}
  {\bibfnamefont {N.~P.}\ \bibnamefont {de~Leon}}, \bibinfo {author}
  {\bibfnamefont {V.}~\bibnamefont {Vuleti{\'c}}}, \ and\ \bibinfo {author}
  {\bibfnamefont {M.~D.}\ \bibnamefont {Lukin}},\ }\href@noop {} {\bibfield
  {journal} {\bibinfo  {journal} {Optica}\ }\textbf {\bibinfo {volume} {2}},\
  \bibinfo {pages} {70} (\bibinfo {year} {2015})}\BibitemShut {NoStop}%
\bibitem [{\citenamefont {Jahnke}\ \emph {et~al.}(2015)\citenamefont {Jahnke},
  \citenamefont {Sipahigil}, \citenamefont {Binder}, \citenamefont {Doherty},
  \citenamefont {Metsch}, \citenamefont {Rogers}, \citenamefont {Manson},
  \citenamefont {Lukin},\ and\ \citenamefont {Jelezko}}]{jahnke2015electron}%
  \BibitemOpen
  \bibfield  {author} {\bibinfo {author} {\bibfnamefont {K.~D.}\ \bibnamefont
  {Jahnke}}, \bibinfo {author} {\bibfnamefont {A.}~\bibnamefont {Sipahigil}},
  \bibinfo {author} {\bibfnamefont {J.~M.}\ \bibnamefont {Binder}}, \bibinfo
  {author} {\bibfnamefont {M.~W.}\ \bibnamefont {Doherty}}, \bibinfo {author}
  {\bibfnamefont {M.}~\bibnamefont {Metsch}}, \bibinfo {author} {\bibfnamefont
  {L.~J.}\ \bibnamefont {Rogers}}, \bibinfo {author} {\bibfnamefont {N.~B.}\
  \bibnamefont {Manson}}, \bibinfo {author} {\bibfnamefont {M.~D.}\
  \bibnamefont {Lukin}}, \ and\ \bibinfo {author} {\bibfnamefont
  {F.}~\bibnamefont {Jelezko}},\ }\href@noop {} {\bibfield  {journal} {\bibinfo
   {journal} {New J. Phys.}\ }\textbf {\bibinfo {volume} {17}},\ \bibinfo
  {pages} {043011} (\bibinfo {year} {2015})}\BibitemShut {NoStop}%
\bibitem [{\citenamefont {Khalid}\ \emph {et~al.}(2015)\citenamefont {Khalid},
  \citenamefont {Chung}, \citenamefont {Rajasekharan}, \citenamefont {Lau},
  \citenamefont {Karle}, \citenamefont {Gibson},\ and\ \citenamefont
  {Tomljenovic-Hanic}}]{khalid2015lifetime}%
  \BibitemOpen
  \bibfield  {author} {\bibinfo {author} {\bibfnamefont {A.}~\bibnamefont
  {Khalid}}, \bibinfo {author} {\bibfnamefont {K.}~\bibnamefont {Chung}},
  \bibinfo {author} {\bibfnamefont {R.}~\bibnamefont {Rajasekharan}}, \bibinfo
  {author} {\bibfnamefont {D.~W.}\ \bibnamefont {Lau}}, \bibinfo {author}
  {\bibfnamefont {T.~J.}\ \bibnamefont {Karle}}, \bibinfo {author}
  {\bibfnamefont {B.~C.}\ \bibnamefont {Gibson}}, \ and\ \bibinfo {author}
  {\bibfnamefont {S.}~\bibnamefont {Tomljenovic-Hanic}},\ }\href@noop {}
  {\bibfield  {journal} {\bibinfo  {journal} {Sci. Rep.}\ }\textbf {\bibinfo
  {volume} {5}},\ \bibinfo {pages} {11179} (\bibinfo {year}
  {2015})}\BibitemShut {NoStop}%
\bibitem [{\citenamefont {Frimmer}\ \emph {et~al.}(2013)\citenamefont
  {Frimmer}, \citenamefont {Mohtashami},\ and\ \citenamefont
  {Femius~Koenderink}}]{frimmer2013nanomechanical}%
  \BibitemOpen
  \bibfield  {author} {\bibinfo {author} {\bibfnamefont {M.}~\bibnamefont
  {Frimmer}}, \bibinfo {author} {\bibfnamefont {A.}~\bibnamefont {Mohtashami}},
  \ and\ \bibinfo {author} {\bibfnamefont {A.}~\bibnamefont
  {Femius~Koenderink}},\ }\href@noop {} {\bibfield  {journal} {\bibinfo
  {journal} {Appl. Phys. Lett.}\ }\textbf {\bibinfo {volume} {102}},\ \bibinfo
  {pages} {121105} (\bibinfo {year} {2013})}\BibitemShut {NoStop}%
\bibitem [{\citenamefont {Patel}\ \emph {et~al.}(2016)\citenamefont {Patel},
  \citenamefont {Schr{\"o}der}, \citenamefont {Wan}, \citenamefont {Li},
  \citenamefont {Mouradian}, \citenamefont {Chen},\ and\ \citenamefont
  {Englund}}]{patel2016efficient}%
  \BibitemOpen
  \bibfield  {author} {\bibinfo {author} {\bibfnamefont {R.~N.}\ \bibnamefont
  {Patel}}, \bibinfo {author} {\bibfnamefont {T.}~\bibnamefont {Schr{\"o}der}},
  \bibinfo {author} {\bibfnamefont {N.}~\bibnamefont {Wan}}, \bibinfo {author}
  {\bibfnamefont {L.}~\bibnamefont {Li}}, \bibinfo {author} {\bibfnamefont
  {S.~L.}\ \bibnamefont {Mouradian}}, \bibinfo {author} {\bibfnamefont {E.~H.}\
  \bibnamefont {Chen}}, \ and\ \bibinfo {author} {\bibfnamefont {D.~R.}\
  \bibnamefont {Englund}},\ }\href@noop {} {\bibfield  {journal} {\bibinfo
  {journal} {Light Sci. Appl.}\ }\textbf {\bibinfo {volume} {5}},\ \bibinfo
  {pages} {e16032} (\bibinfo {year} {2016})}\BibitemShut {NoStop}%
\bibitem [{\citenamefont {Bernien}\ \emph {et~al.}(2013)\citenamefont
  {Bernien}, \citenamefont {Hensen}, \citenamefont {Pfaff}, \citenamefont
  {Koolstra}, \citenamefont {Blok}, \citenamefont {Robledo}, \citenamefont
  {Taminiau}, \citenamefont {Markham}, \citenamefont {Twitchen}, \citenamefont
  {Childress},\ and\ \citenamefont {Hanson}}]{bernien2013heralded}%
  \BibitemOpen
  \bibfield  {author} {\bibinfo {author} {\bibfnamefont {H.}~\bibnamefont
  {Bernien}}, \bibinfo {author} {\bibfnamefont {B.}~\bibnamefont {Hensen}},
  \bibinfo {author} {\bibfnamefont {W.}~\bibnamefont {Pfaff}}, \bibinfo
  {author} {\bibfnamefont {G.}~\bibnamefont {Koolstra}}, \bibinfo {author}
  {\bibfnamefont {M.~S.}\ \bibnamefont {Blok}}, \bibinfo {author}
  {\bibfnamefont {L.}~\bibnamefont {Robledo}}, \bibinfo {author} {\bibfnamefont
  {T.~H.}\ \bibnamefont {Taminiau}}, \bibinfo {author} {\bibfnamefont
  {M.}~\bibnamefont {Markham}}, \bibinfo {author} {\bibfnamefont {D.~J.}\
  \bibnamefont {Twitchen}}, \bibinfo {author} {\bibfnamefont {L.}~\bibnamefont
  {Childress}}, \ and\ \bibinfo {author} {\bibfnamefont {R.}~\bibnamefont
  {Hanson}},\ }\href {\doibase 10.1038/nature12016} {\bibfield  {journal}
  {\bibinfo  {journal} {Nature}\ }\textbf {\bibinfo {volume} {497}},\ \bibinfo
  {pages} {86} (\bibinfo {year} {2013})}\BibitemShut {NoStop}%
\bibitem [{\citenamefont {Goldman}\ \emph {et~al.}(2015)\citenamefont
  {Goldman}, \citenamefont {Sipahigil}, \citenamefont {Doherty}, \citenamefont
  {Yao}, \citenamefont {Bennett}, \citenamefont {Markham}, \citenamefont
  {Twitchen}, \citenamefont {Manson}, \citenamefont {Kubanek},\ and\
  \citenamefont {Lukin}}]{goldman2015phonon}%
  \BibitemOpen
  \bibfield  {author} {\bibinfo {author} {\bibfnamefont {M.~L.}\ \bibnamefont
  {Goldman}}, \bibinfo {author} {\bibfnamefont {A.}~\bibnamefont {Sipahigil}},
  \bibinfo {author} {\bibfnamefont {M.~W.}\ \bibnamefont {Doherty}}, \bibinfo
  {author} {\bibfnamefont {N.~Y.}\ \bibnamefont {Yao}}, \bibinfo {author}
  {\bibfnamefont {S.~D.}\ \bibnamefont {Bennett}}, \bibinfo {author}
  {\bibfnamefont {M.}~\bibnamefont {Markham}}, \bibinfo {author} {\bibfnamefont
  {D.~J.}\ \bibnamefont {Twitchen}}, \bibinfo {author} {\bibfnamefont {N.~B.}\
  \bibnamefont {Manson}}, \bibinfo {author} {\bibfnamefont {A.}~\bibnamefont
  {Kubanek}}, \ and\ \bibinfo {author} {\bibfnamefont {M.~D.}\ \bibnamefont
  {Lukin}},\ }\href@noop {} {\bibfield  {journal} {\bibinfo  {journal} {Phys.
  Rev. Lett.}\ }\textbf {\bibinfo {volume} {114}},\ \bibinfo {pages} {145502}
  (\bibinfo {year} {2015})}\BibitemShut {NoStop}%
\bibitem [{\citenamefont {Schulte}\ \emph {et~al.}(2015)\citenamefont
  {Schulte}, \citenamefont {Hansom}, \citenamefont {Jones}, \citenamefont
  {Matthiesen}, \citenamefont {Le~Gall},\ and\ \citenamefont
  {Atat{\"u}re}}]{schulte2015quadrature}%
  \BibitemOpen
  \bibfield  {author} {\bibinfo {author} {\bibfnamefont {C.~H.}\ \bibnamefont
  {Schulte}}, \bibinfo {author} {\bibfnamefont {J.}~\bibnamefont {Hansom}},
  \bibinfo {author} {\bibfnamefont {A.~E.}\ \bibnamefont {Jones}}, \bibinfo
  {author} {\bibfnamefont {C.}~\bibnamefont {Matthiesen}}, \bibinfo {author}
  {\bibfnamefont {C.}~\bibnamefont {Le~Gall}}, \ and\ \bibinfo {author}
  {\bibfnamefont {M.}~\bibnamefont {Atat{\"u}re}},\ }\href@noop {} {\bibfield
  {journal} {\bibinfo  {journal} {Nature}\ }\textbf {\bibinfo {volume} {525}},\
  \bibinfo {pages} {222} (\bibinfo {year} {2015})}\BibitemShut {NoStop}%
\bibitem [{\citenamefont {Tey}\ \emph {et~al.}(2008)\citenamefont {Tey},
  \citenamefont {Chen}, \citenamefont {Aljunid}, \citenamefont {Chng},
  \citenamefont {Huber}, \citenamefont {Maslennikov},\ and\ \citenamefont
  {Kurtsiefer}}]{tey2008strong}%
  \BibitemOpen
  \bibfield  {author} {\bibinfo {author} {\bibfnamefont {M.~K.}\ \bibnamefont
  {Tey}}, \bibinfo {author} {\bibfnamefont {Z.}~\bibnamefont {Chen}}, \bibinfo
  {author} {\bibfnamefont {S.~A.}\ \bibnamefont {Aljunid}}, \bibinfo {author}
  {\bibfnamefont {B.}~\bibnamefont {Chng}}, \bibinfo {author} {\bibfnamefont
  {F.}~\bibnamefont {Huber}}, \bibinfo {author} {\bibfnamefont
  {G.}~\bibnamefont {Maslennikov}}, \ and\ \bibinfo {author} {\bibfnamefont
  {C.}~\bibnamefont {Kurtsiefer}},\ }\href@noop {} {\bibfield  {journal}
  {\bibinfo  {journal} {Nat. Phys.}\ }\textbf {\bibinfo {volume} {4}},\
  \bibinfo {pages} {924} (\bibinfo {year} {2008})}\BibitemShut {NoStop}%
\bibitem [{\citenamefont {H{\'e}tet}\ \emph {et~al.}(2011)\citenamefont
  {H{\'e}tet}, \citenamefont {Slodi{\v{c}}ka}, \citenamefont {Hennrich},\ and\
  \citenamefont {Blatt}}]{hetet2011single}%
  \BibitemOpen
  \bibfield  {author} {\bibinfo {author} {\bibfnamefont {G.}~\bibnamefont
  {H{\'e}tet}}, \bibinfo {author} {\bibfnamefont {L.}~\bibnamefont
  {Slodi{\v{c}}ka}}, \bibinfo {author} {\bibfnamefont {M.}~\bibnamefont
  {Hennrich}}, \ and\ \bibinfo {author} {\bibfnamefont {R.}~\bibnamefont
  {Blatt}},\ }\href@noop {} {\bibfield  {journal} {\bibinfo  {journal} {Phys.
  Rev. Lett.}\ }\textbf {\bibinfo {volume} {107}},\ \bibinfo {pages} {133002}
  (\bibinfo {year} {2011})}\BibitemShut {NoStop}%
\bibitem [{\citenamefont {Wrigge}\ \emph {et~al.}(2008)\citenamefont {Wrigge},
  \citenamefont {Gerhardt}, \citenamefont {Hwang}, \citenamefont {Zumofen},\
  and\ \citenamefont {Sandoghdar}}]{wrigge2008efficient}%
  \BibitemOpen
  \bibfield  {author} {\bibinfo {author} {\bibfnamefont {G.}~\bibnamefont
  {Wrigge}}, \bibinfo {author} {\bibfnamefont {I.}~\bibnamefont {Gerhardt}},
  \bibinfo {author} {\bibfnamefont {J.}~\bibnamefont {Hwang}}, \bibinfo
  {author} {\bibfnamefont {G.}~\bibnamefont {Zumofen}}, \ and\ \bibinfo
  {author} {\bibfnamefont {V.}~\bibnamefont {Sandoghdar}},\ }\href@noop {}
  {\bibfield  {journal} {\bibinfo  {journal} {Nat. Phys.}\ }\textbf {\bibinfo
  {volume} {4}},\ \bibinfo {pages} {60} (\bibinfo {year} {2008})}\BibitemShut
  {NoStop}%
\bibitem [{\citenamefont {Vamivakas}\ \emph {et~al.}(2007)\citenamefont
  {Vamivakas}, \citenamefont {Atat{\"u}re}, \citenamefont {Dreiser},
  \citenamefont {Yilmaz}, \citenamefont {Badolato}, \citenamefont {Swan},
  \citenamefont {Goldberg}, \citenamefont {Imamoglu},\ and\ \citenamefont
  {{\"U}nl{\"u}}}]{vamivakas2007strong}%
  \BibitemOpen
  \bibfield  {author} {\bibinfo {author} {\bibfnamefont {A.~N.}\ \bibnamefont
  {Vamivakas}}, \bibinfo {author} {\bibfnamefont {M.}~\bibnamefont
  {Atat{\"u}re}}, \bibinfo {author} {\bibfnamefont {J.}~\bibnamefont
  {Dreiser}}, \bibinfo {author} {\bibfnamefont {S.~T.}\ \bibnamefont {Yilmaz}},
  \bibinfo {author} {\bibfnamefont {A.}~\bibnamefont {Badolato}}, \bibinfo
  {author} {\bibfnamefont {A.~K.}\ \bibnamefont {Swan}}, \bibinfo {author}
  {\bibfnamefont {B.~B.}\ \bibnamefont {Goldberg}}, \bibinfo {author}
  {\bibfnamefont {A.}~\bibnamefont {Imamoglu}}, \ and\ \bibinfo {author}
  {\bibfnamefont {M.~S.}\ \bibnamefont {{\"U}nl{\"u}}},\ }\href@noop {}
  {\bibfield  {journal} {\bibinfo  {journal} {Nano Lett.}\ }\textbf {\bibinfo
  {volume} {7}},\ \bibinfo {pages} {2892} (\bibinfo {year} {2007})}\BibitemShut
  {NoStop}%
\end{thebibliography}%

\end{document}